\documentclass[preprint,3p]{elsarticle}

\usepackage{euscript,epsfig,amsmath,amssymb}
\usepackage{amsfonts,latexsym}
\usepackage{array}
\usepackage{enumerate}
\usepackage{booktabs}
\usepackage{tabularx}

\usepackage{hyperref}
\usepackage{url}
\usepackage[normalem]{ulem}
\usepackage{color}

\newcolumntype{L}[1]{>{\hsize=#1\hsize\raggedright\arraybackslash}X}%
\newcolumntype{R}[1]{>{\hsize=#1\hsize\raggedleft\arraybackslash}X}%
\newcolumntype{C}[1]{>{\hsize=#1\hsize\centering\arraybackslash}X}%

\newcommand{\be}[1]{\begin{equation}\label{#1}}
\newcommand{\ee}{\end{equation}}
\newcommand{\ba}[1]{\begin{eqnarray}\label{#1}}
\newcommand{\ea}{\end{eqnarray}}
\newcommand{\rf}[1]{(\ref{#1})}
\newcommand{\nn}{\nonumber}
\renewcommand{\theequation}{\arabic{section}.\arabic{equation}}

\begin{document}

\begin{frontmatter}

\title{Analytic expressions for the second-order scalar perturbations\\ in the $\Lambda$CDM Universe within the cosmic screening approach}

\author[1]{Maxim Eingorn}
\ead{maxim.eingorn@gmail.com}

\author[2]{N. Duygu Guran}
\ead{gurann@itu.edu.tr}

\author[2,3]{Alexander Zhuk}
\ead{ai.zhuk2@gmail.com}

\address[1]{Department of Mathematics and Physics, North Carolina Central University,\\ Fayetteville st. 1801, Durham, North Carolina 27707, U.S.A.}
\address[2]{Department of Physics, Istanbul Technical University, Maslak 34469 Istanbul, Turkey}
\address[3]{Astronomical Observatory, Odessa National University,\\ Dvoryanskaya st. 2, Odessa 65082, Ukraine}

\begin{abstract}
We study the second-order scalar perturbations in the conventional $\Lambda$CDM Universe within the cosmic screening approach. The analytic expressions for both
the velocity-independent and velocity-dependent second-order scalar perturbations are derived. We demonstrate how the Yukawa screening effect, which is inherent
in the first-order metric corrections, manifests itself in the second-order ones. It is shown that the obtained formulas for the second-order perturbations are
reduced to the known post-Newtonian expressions at distances much smaller than the Yukawa screening length. In the era of precision cosmology, these analytic
formulas play an important role since the second-order metric corrections may affect the interpretation of observational data (e.g., the luminosity-redshift
relation, gravitational lensing, baryon acoustic oscillations).
\end{abstract}


\end{frontmatter}



\section{Introduction}
\label{Sec1}

According to the cosmological principle \cite{Weinberg,Ellis}, our Universe is isotropic and homogeneous at large enough scales. This follows from the natural
assumption that the laws of physics should be the same wherever in the Universe we are. Starting from a certain scale, the distribution of inhomogeneities (e.g.,
galaxies and groups of galaxies) should be statistically homogeneous. As a result, such statistically homogeneous Universe has the
Friedmann-Lema$\mathrm{\hat{\i}}$tre-Robertson-Walker (FLRW) metric, and its dynamics is described by the Friedmann equations. This is the zero-order/background
approach. Obviously, inhomogeneities disturb this background resulting in first- and higher-order perturbations. They play a crucial role in the investigation of
the large scale structure formation. Evidently (by definition!), the average values of the first-order perturbations should be equal to zero (see, e.g.,
\cite{EBV}). On the other hand, the average values of the second-order perturbations are different from zero and can affect the background spacetime and matter.
This effect is called backreaction (see, e.g., \cite{BR1,BR2,BR3,BR4,BR5,BR6} and references therein). It is important to determine how strong the backreaction is
and to what extent we may use the standard FLRW metric as a background one. As an example, the backreaction may affect the baryon acoustic oscillations
\cite{BR2,Baumann}. The second-order perturbations also contribute to the luminosity-redshift relation \cite{SecOd1} and gravitational lensing
\cite{SecOd2,SecOd3}.

Within the cosmic screening approach, the theory of the first-order perturbations was developed in the papers \cite{Eingorn-first,MaxRusl1,EKZ1,cosmlaw,EKZ2}. In
its framework, inhomogeneities in the Universe (e.g., galaxies and their groups) are considered as point-like gravitating masses. These masses disturb the
background spacetime and matter. For example, the energy density fluctuation reads $\delta\varepsilon\approx (c^2/a^3)\delta\rho + (3\bar\rho c^2/a^3)\Phi$, where
$c$ denotes the speed of light, $a$ is the scale factor of the Universe, and we singled out the gravitational potential $\Phi$. Consequently, the 00-component of
the Einstein equation is reduced to the Helmholtz-type equation (rather than the Poisson one) for the gravitational potential, where the comoving mass density
fluctuation $\delta\rho =\rho-\bar\rho$ is the source (see Eq.~\rf{2.7} below). As a result, the first-order scalar perturbation $\Phi$ is characterized by
exponential Yukawa screening at large cosmological scales. A similar effect takes place for the first-order vector perturbation ${\bf B}$. The cosmic background,
namely, the nonzero value of the average mass density $\bar\rho$, is responsible for this effect. It is worth mentioning that in our approach we assume that
peculiar velocities of gravitating masses are much less than the speed of light and the gravitational field produced by inhomogeneities is weak (i.e. we work in
the weak field approximation). However, we do not demand the smallness of the mass density contrast $\delta\rho/\overline{\rho}$. Therefore, the given formalism
can be used at all cosmic scales (below and above the Hubble scale).

In the framework of this formalism, the theory of the second-order perturbations was developed in \cite{Eingorn-second}. It is worth noting that the alternative
perturbative method was considered, e.g., in \cite{Adamek}. There, the authors came to the conclusion that the backreaction effects are at a few percent level at
most. In their analysis, in particular, they disregard the term $\sim\Phi^2$. In our present paper, to get the analytic solutions for the second-order
perturbations, we follow the similar method, i.e. we also drop such terms as $\sim\Phi^2$ from the sources of the analyzed perturbations. However, this is the
only similarity with \cite{Adamek}. In other respects our approach is completely different. In \cite{Adamek} the authors actually mix the first- and second-order
perturbations, so their equations are nonlinear. In contrast, in our approach there is no mixture of the orders, so the equations are linear with respect to the
corresponding metric corrections. Thus, in general, we can arrive at different conclusions regarding the backreaction (at least to the extent that may be
important for precision cosmology). Moreover, the interpretation of our results is clearer since the orders are not mixed and it is much easier to deal with
linear equations. As is generally known, from the point of view of numerical calculus, linear equations possess an unquestionable advantage over nonlinear ones,
which consists in higher computational accuracy and saving of computing time.

In \cite{Eingorn-second} the system of equations for the second-order scalar, vector and tensor perturbations was derived. This system has rather complicated
form. However, it is remarkable that these perturbations do not mix, and we can study them separately. In the present paper we study the second-order scalar
perturbations. Even in this case the corresponding equations are still very complicated. So, we restrict ourselves to the analysis of the second-order
perturbations generated by those sources which are dominant at sufficiently small spatial scales, where the mass density contrast is typically large (see also
\cite{Adamek}). This method enables us to solve the reduced equations exactly. In the large-scale spatial regions the linear relativistic perturbation theory
works very well. Hence, the above mentioned simplification is not significant. As a result, we find analytic expressions for the velocity-independent as well as
velocity-dependent scalar perturbations.

\

The paper is structured as follows. In section~\ref{Sec2} we briefly describe the background model and the first-order scalar perturbations. Then, we present the
equations for the second-order scalar perturbations with velocity-independent sources. These equations are solved in section~\ref{Sec3}. The solutions for the
second-order velocity-dependent scalar perturbations are derived in section~\ref{Sec4}. In section~\ref{Sec5} we study the Newtonian and post-Newtonian
approximations for the found analytic solutions. Here we demonstrate that both the velocity-independent and velocity-dependent expressions are reduced to the
known post-Newtonian formulas. In the concluding section~\ref{Sec6} we summarize and discuss the obtained results. In Appendix A we collect formulas which we use
to solve the equations for the second-order scalar perturbations.

\section{From the background model to the second-order scalar perturbations: basic equations}
\label{Sec2}

\setcounter{equation}{0}

\subsection{Background model}

\

We start with the unperturbed FLRW metric
\be{2.1} ds^2=a^2\left(d\eta^2-\delta_{\alpha\beta}dx^{\alpha}dx^{\beta}\right)\, , \quad \alpha,\beta = 1,2,3\, , \ee
where $a(\eta)$ is the scale factor, $\eta$ is the conformal time, and $x^{\alpha}$, $\alpha=1,2,3$, represent the comoving coordinates. It is supposed that the
spatial curvature is zero. Let us write down the corresponding Friedmann equations in the framework of the $\Lambda$CDM model:
\be{2.2}
\frac{3\mathcal{H}^2}{a^2}=\kappa\overline{\varepsilon}+\Lambda\, ,\quad \frac{2\mathcal{H}'+\mathcal{H}^2}{a^2}=\Lambda\, ,
\ee
where $\mathcal{H}\equiv a'/a\equiv (da/d\eta)/a$, with the prime standing for the derivative with respect to $\eta$, while $\kappa\equiv 8\pi G_N/c^4$ (with $c$
denoting the speed of light and $G_N$ being the Newtonian gravitational constant). In addition, $\varepsilon$ stands for the energy density of nonrelativistic
pressureless matter, $\Lambda$ denotes the cosmological constant, and the overline indicates the average. Obviously, the average energy density is defined by the
constant average comoving mass density $\overline{\rho}$ as follows: $\overline{\varepsilon}=\overline{\rho}c^2/a^3$. From Eqs.~\rf{2.2} we can easily get a
useful auxiliary equation
\be{2.3}
\mathcal{H}'-\mathcal{H}^2 = -\frac{\kappa\overline{\rho}c^2}{2a}\, .
\ee

\

\subsection{First-order scalar perturbations}

\

The described above background Universe is perturbed by inhomogeneities in the form of discrete point-like masses with mass density
\be{2.4}
\rho(\eta, \mathbf{r})=\sum_n m_n\delta(\mathbf{r}-\mathbf{r}_n)
\, ,
\ee
where $\mathbf{r}\equiv\left(x^1,x^2,x^3\right)$ is the comoving radius-vector of the observation point, and $\mathbf{r}_n\equiv\left(x_n^1,x_n^2,x_n^3\right)$ is
the comoving radius-vector of the $n$-th particle. These massive particles may represent galaxies and their groups. The mass density fluctuation is
\be{2.5}
\delta \rho = \rho -\overline{\rho}\, .
\ee
It is important to note that we do not assume the smallness of the mass density contrast, i.e. $\delta\rho/\overline{\rho}$ can be much larger than unity. Hence,
our scheme is valid at both superhorizon and subhorizon scales. The inhomogeneities result in scalar perturbations of the metric \rf{2.1}. In the conformal
Newtonian gauge and in the first-order approximation, the perturbed metric is \cite{Bardeen,Mukhanov,Durrer,Rubakov}
\be{2.6} ds^2=a^2\left[\left(1+2\Phi\right)d\eta^2-\left(1-2\Phi\right)\delta_{\alpha\beta}dx^{\alpha}dx^{\beta}\right]\, , \ee
where the first-order scalar perturbation $\Phi (\eta, \mathbf{r})$ satisfies the inequality $|\Phi|\ll 1$. This means that we work in the weak gravitational
field limit. It is well known that, e.g., in the vicinity of galaxies the mass density contrast is much larger than unity, however the gravitational field is
weak. This fact allows expanding the components of the Einstein equation into series in metric corrections. Additionally, we assume that the particle peculiar
velocities are much less than the speed of light: $|\mathbf{v}_n|=|cd\mathbf{r}_n/d\eta|\ll c$. For example, the today's typical values are $(250\div 500)$ km/s.
As demonstrated in \cite{Eingorn-first}, for such nonrelativistic velocities the contribution of the velocity-dependent part into the total expression for the
first-order scalar perturbation $\Phi$ is negligible. In this case the gravitational potential $\Phi$ satisfies the following Helmholtz-type equation
\cite{Eingorn-first}:
\be{2.7}
\Delta\Phi-\frac{a^2}{\lambda^2}\Phi =\frac{\kappa c^2}{2a}\delta\rho\, ,
\ee
where $\Delta\equiv\delta^{\alpha\beta}\frac{\partial^2}{\partial x^{\alpha}\partial x^{\beta}}$ is the Laplace operator in comoving coordinates. The
time-dependent parameter
\be{2.8}
\lambda\equiv \left[\frac{3\kappa\overline{\rho}c^2}{2a^3}\right]^{-1/2}
\ee
defines the characteristic Yukawa screening length of the gravitational interaction.

The solution of Eq.~\rf{2.7} is \cite{Eingorn-first}
\be{2.9} \Phi(\eta,\mathbf{r})= \frac13 -\frac{\kappa c^2}{8\pi a}\sum_n\frac{m_n}{|\mathbf{r}-\mathbf{r}_n|}e^{-\mu|\mathbf{r}-\mathbf{r}_n|}\, , \ee
where
\be{2.10}
\mu=\frac{a}{\lambda} = \left[\frac{3\kappa \overline{\rho} c^2}{2 a}\right]^{1/2}\, .
\ee

\

\subsection{Second-order velocity-independent scalar perturbations}

\

Let us turn now to the second-order scalar perturbations $\Phi^{(2)}(\eta,\mathbf{r})$ and $\Psi^{(2)}(\eta,\mathbf{r})$. The corresponding metric reads
\be{2.11} ds^2=a^2\left[\left(1+2\Phi +2\Phi^{(2)}\right)d\eta^2-\left(1-2\Phi-2\Psi^{(2)}\right)\delta_{\alpha\beta}dx^{\alpha}dx^{\beta}\right]\, . \ee

The main aim of the present paper consists in determination of these perturbations. According to \cite{Eingorn-second}, the functions
$\Phi^{(2)}(\eta,\mathbf{r})$ and $\Psi^{(2)}(\eta,\mathbf{r})$ satisfy the following system of master equations:
\ba{2.12}
\Delta\Psi^{(2)}-\frac{a^2}{\lambda^2}\Psi^{(2)}&=&\frac{a^2}{2}Q_{00}+\frac{3a^2}{2}\mathcal{H}Q^{(\parallel)}\, ,\\
\label{2.13} \Phi^{(2)}-\Psi^{(2)}&=&a^2Q^{(S)}\, , \ea
where we momentarily keep only the velocity-independent parts of the sources:
\ba{2.14}
Q_{00}&=&-\left(\frac{3\kappa\bar{\rho}c^2}{2a^3}+\frac{15}{a^2}\mathcal{H}^2\right)\Phi^2
-\frac{2}{a^2}\Phi\Delta\Phi-\frac{3}{a^2}(\nabla\Phi)^2\, ,\\
\label{2.15}
\Delta Q^{(\parallel)}&=& \Delta \left(\frac{5}{a^2}\mathcal{H}\Phi^2\right)\, ,\\
\label{2.16} \Delta \Delta Q^{(S)}&=&-\frac{1}{2}\Delta Q_{\alpha \alpha}+\frac{3}{2}\frac{\partial^2Q_{\alpha \beta}}{\partial x^\alpha \partial x^\beta}\, , \ea
\ba{2.17} Q_{\alpha \alpha}&\equiv& Q_{11}+Q_{22}+Q_{33}
=\left(\frac{12\kappa \bar{\rho}c^2}{a^3}-\frac{15}{a^2}\mathcal{H}^2 \right)\Phi^2-\frac{8}{a^2}\Phi \Delta \Phi-\frac{7}{a^2}(\nabla \Phi)^2\, ,\\
\label{2.18}
Q_{\alpha \beta}&=&\frac{2}{a^2}\frac{\partial \Phi}{\partial x^\alpha}\frac{\partial \Phi}{\partial x^\beta}+\frac{4}{a^2}
\Phi\frac{\partial^2\Phi}{\partial x^\alpha\partial x^\beta}\, ,\quad \alpha\neq\beta\, ,\\
\label{2.19}
Q_{\beta \beta}&=&\left(\frac{4\kappa \bar{\rho}c^2}{a^3}-\frac{5}{a^2}\mathcal{H}^2 \right)\Phi^2-\frac{4}{a^2}\Phi \Delta \Phi-\frac{3}{a^2}(\nabla \Phi)^2\nn \\
&+&\frac{4}{a^2}\Phi\frac{\partial^2 \Phi}{\partial x^{\beta^2}}+\frac{2}{a^2} \left(\frac{\partial \Phi}{\partial x^\beta}\right)^2\, ,\quad \beta=1,2,3\, , \ea
where $(\nabla \Phi)^2 \equiv \nabla \Phi \nabla \Phi= \delta^{\alpha\beta}\frac{\partial\Phi}{\partial x^{\alpha}}\frac{\partial\Phi}{\partial x^{\beta}}$. These
expressions follow from Eqs.~(3.30), (3.45), (3.34), (3.42), (3.35), (3.33) and (3.32) in \cite{Eingorn-second} where we disregard the peculiar velocities and
vector perturbations. Now, we should solve these equations.

\section{Analytic solutions for the second-order velocity-independent potentials $\Phi^{(2)}$ and $\Psi^{(2)}$}
\label{Sec3}

\setcounter{equation}{0}

\subsection{Potential $\Psi^{(2)}(\eta,\mathbf{r})$}

\

The potential $\Psi^{(2)}$ is the solution of Eq.~\rf{2.12}. To solve this equation, we should determine its right-hand side. First, from Eq.~\rf{2.15} we obtain
the function $Q^{(\parallel)}$:
\be{3.1}
 Q^{(\parallel)}=\frac{5}{a^2}\mathcal{H}\Phi^2\, .
\ee
Taking into account Eq.~\rf{2.7} for $\Delta \Phi$, the function $Q_{00}$ \rf{2.14} can be rewritten in the form
\be{3.2}
Q_{00}=-\left(\frac{9\kappa\bar{\rho}c^2}{2a^3}+\frac{15}{a^2}\mathcal{H}^2\right)\Phi^2-\frac{\kappa c^2}{a^3}\Phi\delta\rho-\frac{3}{a^2}(\nabla\Phi)^2\, .
\ee
Then, Eq.~\rf{2.12} takes the form
\ba{3.3}
&{}&\Delta\Psi^{(2)}-\frac{3\kappa\bar{\rho}c^2}{2a}\Psi^{(2)}=-\frac{9\kappa\bar{\rho}c^2}{4a}\Phi^2-
\frac{\kappa c^2}{2a}\Phi\delta\rho-\frac{3}{2}(\nabla\Phi)^2\nn\\
&=&-\frac{9\kappa\bar{\rho}c^2}{4a}\Phi^2-\frac{\kappa c^2}{2a}\Phi\delta\rho-\frac{3}{4}\Delta(\Phi^2)+\frac{3}{2}\Phi\Delta\Phi
=-\frac{3}{4}\Delta(\Phi^2)+\frac{\kappa c^2}{4a}\Phi\delta\rho\, ,
\ea
where we used Eq.~\rf{2.7} and the auxiliary equality
\be{a.6}(\nabla \Phi)^2=\frac{1}{2}\Delta\left(\Phi ^2\right)-\Phi \Delta \Phi\, .\ee

It makes sense to define a new function
\be{3.4}
\chi =\Psi^{(2)}+\frac{3}{4}\Phi^2\, ,
\ee
which satisfies the equation
\be{3.5} \Delta \chi-\frac{3\kappa\bar{\rho}c^2}{2a}\chi=-\frac{9\kappa\bar{\rho}c^2}{8a}\Phi^2+\frac{\kappa c^2}{4a}\Phi\delta\rho\, . \ee
To solve this equation analytically, we resort to a supplementary simplification. Namely, we concentrate on those sources of the second-order perturbations, which
dominate at sufficiently small distances where the mass density contrast is typically large. In the case of Eq.~\rf{3.5} this means dropping the term $\sim\Phi^2$
while keeping the term $\sim\Phi\delta\rho$ in the right-hand side. Such a simplification implies failing to take into account all sources at large enough
distances, but this failure is insignificant since the disregarded sources (such as the term $\sim\Phi^2$) are much less than the corresponding first-order ones
in the considered large-scale spatial region, where the linear relativistic perturbation theory works very well (see the argumentation in \cite{Adamek} and Refs.
therein). Returning to Eq.~\rf{3.5}, we have
\be{3.6} \Delta \chi-\frac{3\kappa\bar{\rho}c^2}{2a}\chi=\frac{\kappa c^2}{4a}\Phi\delta\rho\, . \ee
This is the Helmholtz-type equation with the Green's function
\be{3.7} G_{\mathrm{H}}(\text{\textbf{r}},\text{\textbf{r}}') =-\frac{1}{4\pi}\frac{
e^{-a|\text{\textbf{r}}-\text{\textbf{r}}'|/\lambda}}{|\text{\textbf{r}}-\text{\textbf{r}}'|}\, . \ee
Therefore, we look for its solution in the form
\ba{3.8} \chi&=& \int d \text{\textbf{r}}' \left( -\frac{1}{4\pi}\frac{
e^{-a|\text{\textbf{r}}-\text{\textbf{r}}'|/\lambda}}{|\text{\textbf{r}}-\text{\textbf{r}}'|}
\right)\left(\frac{\kappa c^2}{4a}\Phi(\eta,\text{\textbf{r}}')\delta\rho(\eta,\text{\textbf{r}}')\right)\nn\\
&=&-\frac{1}{4\pi}\frac{\kappa c^2}{4a}\int d \text{\textbf{r}}' \frac{ e^{-a|\text{\textbf{r}}-\text{\textbf{r}}'|/\lambda}}{|\text{\textbf{r}}-
\text{\textbf{r}}'|}\left(\frac{1}{3}-\frac{\kappa c^2}{8\pi a}\sum\limits_{k} \frac{m_k e^{-a|\text{\textbf{r}}'-
\text{\textbf{r}}_k|/\lambda}}{|\text{\textbf{r}}'-\text{\textbf{r}}_k|}\right)\left(\sum\limits_{k'} m_{k'}\delta(\text{\textbf{r}}'-
\text{\textbf{r}}_{k'})-\bar{\rho}\right)\nn\\
&=& -\frac{\kappa c^2}{16\pi a}\left\{-\frac{\bar{\rho}}{3}I_1+\frac{1}{3}\sum\limits_{k} m_{k}\ \frac{ e^{-a|\text{\textbf{r}}-
\text{\textbf{r}}_k|/\lambda}}{|\text{\textbf{r}}-\text{\textbf{r}}_k|}-\frac{\kappa c^2}{8\pi a}{\sum\limits_{k,k'}}^{'}m_k m_{k'}\ \frac{
e^{-a|\text{\textbf{r}}-\text{\textbf{r}}_{k'} |/\lambda}}{|\text{\textbf{r}}-\text{\textbf{r}}_{k'}|}
\frac{e^{-a|\text{\textbf{r}}_{k'}-\text{\textbf{r}}_k|/\lambda}}{|\text{\textbf{r}}_{k'}-\text{\textbf{r}}_k|}\right.\nn\\
&+&\left. \frac{\kappa \bar{\rho} c^2}{8\pi a}\sum\limits_{k}m_k\ I_2\right\}\, ,
\ea
where the integrals $I_1$ and $I_2$ are given by the formulas \rf{b.1} and \rf{b.2}, respectively. The prime over the double sum indicates that the summation
indices must not coincide. Substituting these integrals and taking into account Eq.~\rf{3.4}, we finally get
\ba{3.9}
\Psi^{(2)}&=& -\frac{3}{4}\Phi^2+\chi=-\frac{3}{4}\Phi^2+\frac{\Phi}{6}-\frac{\pi \overline{\rho} \lambda}{ a}
\left(\frac{\kappa c^2}{8\pi a}\right)^2\sum\limits_{k}m_k\ e^{-a|\text{\textbf{r}}-\text{\textbf{r}}_k|/\lambda}\nonumber \nn\\
&+&\frac{1}{2}\left(\frac{\kappa c^2}{8\pi a}\right)^2{\sum\limits_{k,k'}}^{'}m_k m_{k'}\ \frac{
e^{-a|\text{\textbf{r}}-\text{\textbf{r}}_{k}|/\lambda}}{|\text{\textbf{r}}-\text{\textbf{r}}_{k}|}
\frac{e^{-a|\text{\textbf{r}}_{k'}-\text{\textbf{r}}_k|/\lambda}}{|\text{\textbf{r}}_{k'}-\text{\textbf{r}}_k|}\, . \ea
In spite of the presence of the first-order term in the right-hand side, this function in total is of the second order. We clearly demonstrate it for the case of
the small-scale limit in section~\ref{Sec5} (see the formula \rf{4.8}).

\

\subsection{Potential $\Phi^{(2)}(\eta,\mathbf{r})$}

\

According to Eq.~\rf{2.13}, to get the potential $\Phi^{(2)}$, we need to determine the function $Q^{(S)}$. This function satisfies Eq.~\rf{2.16} where
$Q_{\alpha\alpha}$ and $Q_{\alpha\beta}$ are defined by Eqs.~\rf{2.17}-\rf{2.19}. The function  $Q_{\alpha\alpha}$ can be rewritten as follows:
\be{3.10}
Q_{\alpha \alpha}=-\frac{15}{a^2}\mathcal{H}^2\Phi^2-\frac{4\kappa c^2}{a^3}\Phi\delta\rho-\frac{7}{a^2}(\nabla\Phi)^2\, ,
\ee
where we used Eq.~\rf{2.7}. After lengthy calculations one can also derive
\be{3.12} \frac{\partial^2Q_{\alpha \beta}}{\partial x^\alpha \partial x^\beta}=-\frac{\kappa c^2}{a^3}\nabla\left(\delta\rho\nabla\Phi\right)-
\frac{5}{a^2}\mathcal{H}'\Delta(\Phi^2)\, . \ee
Therefore, the function $Q^{(S)}$ satisfies the following equation:
\be{3.13} \Delta \Delta Q^{(S)}=-\frac{1}{2}\Delta Q_{\alpha \alpha}-\frac{15}{2a^2}\mathcal{H}'\Delta(\Phi^2)+\frac{3 \kappa \bar{\rho} c^2}{2 a^3}\Delta\Phi-
\frac{3 \kappa c^2}{2 a^3}\nabla\left(\rho\nabla\Phi\right)\, . \ee

To solve this equation, we introduce a new function
\be{3.17} f=\frac{1}{4\pi}\sum_{k}m_k\frac{1}{|\text{\textbf{r}}-\text{\textbf{r}}_k|^3}(\text{\textbf{r}}- \text{\textbf{r}}_k)\nabla
\Phi(\eta,\text{\textbf{r}}_k)\, , \ee
which satisfies the equation
\be{3.14} \Delta f=\nabla\left(\rho\nabla\Phi\right)\, . \ee

Now, applying the inverse Laplace operator $\Delta^{-1}$ to Eq.~\rf{3.13}, we get
\ba{3.18}
\Delta Q^{(S)}&=&-\frac{1}{2} Q_{\alpha \alpha}-\frac{15}{2a^2}\mathcal{H}'\Phi^2+\frac{3 \kappa \bar{\rho} c^2}{2 a^3}\Phi-\frac{3 \kappa c^2}{2 a^3} f\nn \\
&=& \frac{15\kappa \bar{\rho} c^2}{4a^3}\Phi^2+\frac{2\kappa c^2}{a^3}\Phi\rho-\frac{ \kappa \bar{\rho} c^2}{2 a^3}\Phi+\frac{7}{2a^2}(\nabla\Phi)^2\nn \\
&-&\frac{3 \kappa c^2}{8\pi a^3} \sum_{k}m_k\frac{1}{|\text{\textbf{r}}-\text{\textbf{r}}_k|^3}(\text{\textbf{r}}- \text{\textbf{r}}_k)\nabla
\Phi(\eta,\text{\textbf{r}}_k)\, , \ea
where the function $Q_{\alpha\alpha}$ is given by \rf{3.10}. With the help of Eqs.~\rf{2.7} and \rf{a.6} as well as the formula
$1/{|\text{\textbf{r}}-\text{\textbf{r}}_k|}=\frac{1}{2}\Delta|\text{\textbf{r}}-\text{\textbf{r}}_k|$ (see, e.g., $\S$ 106 in \cite{Landau}) we finally arrive at
the equation
\be{3.19}
\Delta Q^{(S)}
    =\frac{7}{4a^2}\Delta\left(\Phi ^2\right)+\frac{3 \kappa c^2}{16\pi a^3} \sum_{k}m_k\Delta
    \left(\frac{(\text{\textbf{r}}-\text{\textbf{r}}_k)\nabla \Phi(\eta,\text{\textbf{r}}_k)}{|\text{\textbf{r}}-\text{\textbf{r}}_k|}\right)+
    \frac{\kappa c^2}{4a^3}\Phi\rho+\frac{ 5\kappa \bar{\rho} c^2}{4 a^3}\Phi-\frac{3\kappa \bar{\rho} c^2}{2a^3}\Phi^2\, .
\ee
Using the same reasoning as for Eq.~\rf{3.5}, we drop the term $\sim\Phi^2$ in the right-hand side. The solution of the resulting equation can be found with the
help of the Laplace operator Green's function
\be{3.16} G_{\mathrm{L}}(\text{\textbf{r}},\text{\textbf{r}}')=-\frac{1}{4\pi}\frac{1}{|\text{\textbf{r}}-\text{\textbf{r}}'|}\, . \ee
Let us introduce two new functions $F_1$ and $F_2$:
\ba{3.20} \Delta F_1=\Phi\rho &\implies& F_1=\int d\text{\textbf{r}}' G_{\mathrm{L}}(\text{\textbf{r}},\text{\textbf{r}}')
\Phi(\eta,\text{\textbf{r}}')\rho(\eta,\text{\textbf{r}}')\, ,\\
\Delta F_2=\Phi &\implies& F_2=\int d\text{\textbf{r}}' G_{\mathrm{L}}(\text{\textbf{r}},\text{\textbf{r}}') \Phi(\eta,\text{\textbf{r}}')\, . \label{3.21}\ea
Therefore, for $Q^{(S)}$ we get
\be{3.22} Q^{(S)} =\frac{7}{4a^2}\Phi ^2+\frac{3 \kappa c^2}{16\pi a^3} \sum_{k}m_k \left(\frac{(\text{\textbf{r}}-\text{\textbf{r}}_k) \nabla
\Phi(\eta,\text{\textbf{r}}_k)}{|\text{\textbf{r}}-\text{\textbf{r}}_k|}\right)+\frac{\kappa c^2}{4a^3} F_1+\frac{ 5\kappa \bar{\rho} c^2}{4 a^3} F_2\, , \ee
where
\be{3.23} F_1 =-\frac{1}{4\pi}\sum\limits_{k} \frac{ m_k}{|\text{\textbf{r}}-\text{\textbf{r}}_k|}\left(\frac{1}{3}-\frac{\kappa c^2}{8\pi a} \sum\limits_{l}
m_l\frac{e^{ -\mu |\text{\textbf{r}}_k-\text{\textbf{r}}_l|}}{|\text{\textbf{r}}_k-\text{\textbf{r}}_l|}\right) =-\frac{1}{4\pi}\sum\limits_{k} \frac{
m_k}{|\text{\textbf{r}}-\text{\textbf{r}}_k|}\Phi(\eta,\text{\textbf{r}}_k) \ee
and
\be{3.24}
F_2
=-\frac{1}{4\pi}\int\frac{ d\text{\textbf{r}}'}{|\text{\textbf{r}}-\text{\textbf{r}}'|}\left(\frac{1}{3}-\frac{\kappa c^2}{8\pi a}
\sum\limits_{k} \frac{m_k}{|\text{\textbf{r}}'-\text{\textbf{r}}_k|}e^{ -\mu |\text{\textbf{r}}'-\text{\textbf{r}}_k|}\right)\, .
\ee
The function $F_2$ can be also expressed as follows:
\be{3.25} F_2=\frac{1}{12\pi\overline{\rho}}\left[\int d\text{\textbf{r}}'\frac{\delta\rho (\eta,\text{\textbf{r}}') }{|\text{\textbf{r}}-\text{\textbf{r}}'|}
-\sum\limits_{k}m_k\frac{e^{ -\mu |\text{\textbf{r}}-\text{\textbf{r}}_k|}}{|\text{\textbf{r}}-\text{\textbf{r}}_k|}\right]\, , \ee
where we used the integral \rf{b.3}.

Finally, the potential $\Phi^{(2)}$ reads:
\be{3.26}
\Phi^{(2)}=\Psi^{(2)} +a^2 Q^{(S)}\, ,
\ee
where $\Psi^{(2)}$ and $Q^{(S)}$ are given by Eqs.~\rf{3.9} and \rf{3.22}, respectively.


\section{Second-order velocity-dependent potentials $\Phi^{(2)}_{(v)}$ and $\Psi^{(2)}_{(v)}$}
\label{Sec4}

\setcounter{equation}{0}

Let us turn now to the analytic expressions for the second-order velocity-dependent potentials. In the framework of the cosmic screening approach, the
velocity-independent and velocity-dependent sources of the potentials enter into the corresponding equations additively. For example, the right-hand sides of
Eqs.~\rf{2.12} and \rf{2.13} (Eqs.~(3.56) in \cite{Eingorn-second}) clearly demonstrate it. Therefore, one can find the respective solutions separately. The
velocity-independent parts of the second-order potentials were obtained in the previous section. To derive these parts, we focused on the gravitational field
sources prevailing in small enough regions with considerable density contrasts. For these regions the velocity-dependent sources of the second-order perturbations
are substantially simplified as well (see section~3.4 in \cite{Eingorn-second}, devoted to comparison of quantities of different orders, and also
\cite{Baumann,Adamek} for the similar reasoning). For instance, the term $\sim \rho{\bf\tilde v}{\bf B}$ (the last one in the total expression for $Q_{00}$, given
by Eq.~(3.30) in \cite{Eingorn-second}) is negligible as compared to the term $\sim \rho\tilde v^2$ (the first one in the same formula) in view of the fact that
${\bf B}$ is of the order of ${\bf \tilde v}\Phi$ at small scales. Consequently, the ratio $\left(\rho{\bf\tilde v}{\bf B}\right) / \left(\rho\tilde v^2\right)$
is of the order of $|\Phi|\ll1$. Adhering to the considered supplementary simplification, from Eqs.~(3.30)-(3.35) in \cite{Eingorn-second} we get
\begin{flalign}
Q_{00}^{(v)}&=\frac{\kappa c^2}{2a^3}\rho \tilde{v}^2\, , \label{5.1}\\
Q_{0\beta}^{(v)}&=0\, ,\quad \beta=1,2,3\, , \label{5.2}\\
Q_{\alpha\beta}^{(v)}&=-\frac{\kappa c^2}{a^3}\rho \tilde{v}^{\alpha} \tilde{v}^{\beta}\, ,\quad \alpha,\beta =1,2,3\, ,\label{5.4}\\
\frac{\partial Q_{0 \beta}^{(v)}}{\partial x^\beta}&\equiv\frac{\partial Q_{01}^{(v)}}{\partial x^1}+
\frac{\partial Q_{02}^{(v)}}{\partial x^2}+\frac{\partial Q_{03}^{(v)}}{\partial x^3}=0\, , \label{5.5}\\
Q_{\alpha \alpha}^{(v)}&\equiv Q_{11}^{(v)}+Q_{22}^{(v)}+Q_{33}^{(v)}=-\frac{\kappa c^2}{a^3}\rho \tilde{v}^2\, , \label{5.6}
\end{flalign}
where $\tilde v^{\beta}\equiv dx^{\beta}/d\eta$ is connected with the physical peculiar velocity $ v^{\beta}\equiv a dx^{\beta}/dt$ as follows: $v^{\beta}=c\tilde
v^{\beta}$ (in view of the relationship $ad\eta=cdt$ defining the synchronous time $t$). In addition, $\tilde v^2\equiv\delta_{\alpha\beta}\tilde v^{\alpha}\tilde
v^{\beta}$. Here and in what follows, both the superscript ``$(v)$'' and the subscript ``$(v)$'' indicate that the corresponding quantity depends on velocities of
massive particles. The function $Q^{(\parallel)}$ is defined by Eq.~(3.45) in \cite{Eingorn-second}. Taking into account \rf{5.5}, we get
\be{5.7} Q^{(\parallel)}_{(v)}=0\, . \ee
The function $Q^{(S)}$ satisfies Eq.~\rf{2.16} where $Q_{\alpha\beta}$ and $Q_{\alpha\alpha}$ are given now by \rf{5.4} and \rf{5.6}, respectively, with the mass
density \rf{2.4}. Then, the solution of \rf{2.16} can be written in the form
\be{5.8} Q^{(S)}_{(v)}=\frac{\kappa c^2}{2a^3}\sum\limits_{n} m_n \tilde{v}_n^2 \Delta ^{-1}\left[\delta({\bf r}-{\bf r}_{n}) \right]-\frac{3\kappa
c^2}{2a^3}\frac{\partial ^2}{\partial x^\alpha \partial x^\beta} \sum\limits_{n} m_n \tilde{v}_n^\alpha \tilde{v}_n^\beta \Delta ^{-1}\Delta
^{-1}\left[\delta({\bf r}-{\bf r}_{n}) \right]\, . \ee
The inverse Laplacians in this equation are easily solvable:
\be{5.9}
 \Delta ^{-1}\left[\delta({\bf r}-{\bf r}_{n}) \right]=-\frac{1}{4\pi}\frac{1}{|{\bf r}-{\bf r}_{n}|}\, ,\quad
\Delta ^{-1}\Delta ^{-1}\left[\delta({\bf r}-{\bf r}_{n}) \right]=-\frac{1}{8\pi}|{\bf r}-{\bf r}_{n}|\, . \ee
Therefore,
\be{5.10} a^2 Q^{(S)}_{(v)}=-\frac{\kappa c^2}{8 \pi a}\sum\limits_{n}\frac{m_n \tilde{v}_n^2}{|{\bf r}-{\bf r}_{n}|}+\frac{3\kappa c^2}{16 \pi
a}\sum\limits_{n}m_n\tilde{v}_n^\alpha \tilde{v}_n^\beta \frac{\partial ^2}{\partial x^\alpha \partial x^\beta} |{\bf r}-{\bf r}_{n}|\, . \ee
Now we can derive expressions for the velocity-dependent second-order potentials. The potential $\Psi^{(2)}$ satisfies Eq.~\rf{2.12} where we should substitute
the functions \rf{5.1} and \rf{5.7}. Then, this equation reads
\be{5.11} \Delta  \Psi^{(2)}_{(v)}-\frac{3\kappa\bar{\rho}c^2}{2a}\Psi^{(2)}_{(v)}=\frac{\kappa c^2}{4a} \sum\limits_{n} m_n \tilde{v}_n^2 \delta({\bf r}-{\bf
r}_{n})\, .\ee
It has the following solution:
\be{5.12} \Psi^{(2)}_{(v)}=-\frac{\kappa c^2}{16 \pi a}\sum\limits_{n}\frac{m_n \tilde{v}_n^2}{|{\bf r}-{\bf r}_{n}|}e^{-\mu |{\bf r}-{\bf r}_{n}|}\, , \ee
where $\mu$ is given by \rf{2.10}. 
The potential $\Phi^{(2)}$ satisfies Eq.~\rf{2.13} where we should substitute \rf{5.10} and \rf{5.12}:
\be{5.13} \Phi^{(2)}_{(v)}= -\frac{\kappa c^2}{16 \pi a}\sum\limits_{n}\frac{m_n \tilde{v}_n^2}{|{\bf r}-{\bf r}_{n}|}e^{-\mu |{\bf r}-{\bf r}_{n}|} -\frac{\kappa
c^2}{8 \pi a}\sum\limits_{n}\frac{m_n \tilde{v}_n^2}{|{\bf r}-{\bf r}_{n}|}+\frac{3\kappa c^2}{16 \pi a}\sum\limits_{n}m_n\tilde{v}_n^\alpha \tilde{v}_n^\beta
\frac{\partial ^2}{\partial x^\alpha \partial x^\beta} |{\bf r}-{\bf r}_{n}|\, . \ee


\section{Newtonian and  post-Newtonian cosmological approximations}
\label{Sec5}

\setcounter{equation}{0}

In the present section we consider the derived above formulas at distances much smaller than the screening length: $\mu r = a r/\lambda =r_{\mathrm{ph}}/\lambda
\ll 1$, where $r_{\mathrm{ph}} =ar$ is the physical distance.
Thereby we analyze the Newtonian and post-Newtonian approximations. We call them cosmological since the obtained expressions depend on the scale factor $a$. To
study these limits, we consider an auxiliary model. In this model a sphere of comoving radius $R$ contains $N$ discrete particles. Outside this sphere, the rest
of the Universe is uniformly filled with matter with the constant comoving mass density $\overline{\rho}$. For such a geometrical configuration, we will get the
first- and second-order perturbations in a point with radius-vector ${\bf r}$ inside the sphere: $r<R$. Additionally, we assume that the physical radius of the
sphere is much less than the screening length: $aR/\lambda=\mu R \ll 1$.

\

\subsection{Gravitational potential}

\

It is well known (see, e.g., \cite{Landau}) that the first-order perturbation $\Phi(\eta,\mathbf{r})$ corresponds to the gravitational potential. In the cosmic
screening approach this function is given by Eq.~\rf{2.9}. In the case of the described above model, the function \rf{2.9} takes the form
\be{4.1} \Phi =\frac{1}{3}-\frac{\kappa c^2}{8\pi a}\sum\limits_{n=1}^N \frac{m_n}{|\text{\textbf{r}}-
\text{\textbf{r}}_n|}e^{-\mu|\text{\textbf{r}}-\text{\textbf{r}}_n|}- \frac{\kappa \bar{\rho} c^2}{8\pi a}\int\limits_{r'>R} d{\bf
r}'\frac{e^{-\mu|\text{\textbf{r}}-\text{\textbf{r}}'|}}{|\text{\textbf{r}}-\text{\textbf{r}}'|}\, , \ee
where the sum is taken over all discrete masses inside the sphere (i.e. $r_n<R$).
Since we consider the point inside the sphere ($r<R$), the integral in \rf{4.1} coincides with $\mathcal{I}_1(r<R)$ given by \rf{b.6}. Therefore,
\be{4.2} \Phi =\frac{1}{3}-\frac{\kappa c^2}{8\pi a}\sum\limits_{n=1}^N \frac{m_n}{|\text{\textbf{r}}-\text{\textbf{r}}_n|}e^{-\mu|\text{\textbf{r}}-
\text{\textbf{r}}_n|}-\frac{1}{3}e^{-\mu R}\left(1+\mu R\right)\frac{\sinh (\mu r)}{\mu r}\, . \ee
In the limit $\mu R \to 0 \ \Rightarrow\ \mu r \to 0,\ \mu r_n \to 0$ we obtain
\be{4.3} \Phi \approx -\frac{\kappa c^2}{8\pi a}\sum\limits_{n=1}^N \frac{m_n}{|\text{\textbf{r}}-\text{\textbf{r}}_n|} \equiv\Phi_N\, , \ee
where $\Phi_N$ is the Newtonian potential at the position $\mathbf{r}$ inside the sphere, produced by all $N$ discrete masses. It is worth reminding that $\kappa
c^2/(8\pi) =G_N/c^2$ and the physical distance $r_{\mathrm{ph}}=a r$. The formula \rf{4.3} clearly demonstrates that for the considered model and in the given
approximation the gravitational potential is determined by the particles from the nearest environment, and the term $1/3$ is exactly compensated by the
contribution of an infinite number of remote particles. As a result, the discussed expression is truly of the first order of smallness.

\

\subsection{Velocity-independent second-order perturbations}

\

Let us study now 
the second-order perturbation $\Psi^{(2)}(\eta,\mathbf{r})$ given by the formula \rf{3.9}. According to the previous subsection, in the Newtonian limit $\Phi^2
\to \Phi_N^2$. Therefore, we only need to investigate the function $\chi$. For the considered model with $N$ discrete particles inside the sphere and uniformly
distributed matter outside the sphere, the function $\chi$ takes the form
\ba{4.4} &&\chi=\frac{1}{18}-\frac{\mu^2}{72\pi \bar{\rho}}\sum\limits_{k=1}^{N} m_{k}\ \frac{ e^{-\mu |\text{\textbf{r}}-
\text{\textbf{r}}_k|}}{|\text{\textbf{r}}-\text{\textbf{r}}_k|}-\frac{\mu^3}{144\pi \bar{\rho}}\sum\limits_{k=1}^{N}m_k\ e^{-\mu |\text{\textbf{r}}-
\text{\textbf{r}}_k|}\nn\\
&+&\frac{\mu^4}{288\pi^2 \bar{\rho}^2}{\sum\limits_{k,k'=1}^{N}} {\vphantom{\int}}' m_k m_{k'}\ \frac{ e^{-\mu |\text{\textbf{r}}-\text{\textbf{r}}_{k}
|}}{|\text{\textbf{r}}- \text{\textbf{r}}_{k}|} \frac{e^{-\mu|\text{\textbf{r}}_{k'}-\text{\textbf{r}}_k|}}{|\text{\textbf{r}}_{k'}-\text{\textbf{r}}_k|}-
\frac{\mu^2}{72\pi}\int\limits_{r'>R} d{\bf r}'\ \frac{ e^{-\mu|\text{\textbf{r}}-\text{\textbf{r}}'|}}{|\text{\textbf{r}}-\text{\textbf{r}}'|}\nn\\
&-&\frac{\mu^3}{144\pi }\int\limits_{r'>R} d{\bf r}'\ e^{-\mu|\text{\textbf{r}}-\text{\textbf{r}}'|}+ \frac{\mu^4}{288\pi^2 }\int\limits_{r'>R} d{\bf r}'
\int\limits_{r''>R} d{\bf r}''\ \frac{ e^{-\mu|\text{\textbf{r}}-\text{\textbf{r}}' |}}{|\text{\textbf{r}}-
\text{\textbf{r}}'|} \frac{e^{-\mu |\text{\textbf{r}}''-\text{\textbf{r}}'|}}{|\text{\textbf{r}}''-\text{\textbf{r}}'|}\nn\\
&+&\frac{\mu^4}{288\pi^2 \bar{\rho}}\sum\limits_{k=1}^{N}m_k\frac{ e^{-\mu |\text{\textbf{r}}-\text{\textbf{r}}_{k} |}}{|\text{\textbf{r}}-
\text{\textbf{r}}_{k}|}\ \int\limits_{r'>R} d{\bf r}'\  \frac{e^{-\mu|\text{\textbf{r}}'- \text{\textbf{r}}_k|}}{|\text{\textbf{r}}'-\text{\textbf{r}}_k|}
+\frac{\mu^4}{288\pi^2 \bar{\rho}}\sum\limits_{k=1}^{N}m_{k}\int\limits_{r'>R} d{\bf r}' \ \frac{ e^{-\mu |\text{\textbf{r}}- \text{\textbf{r}}'
|}}{|\text{\textbf{r}}-\text{\textbf{r}}'|}\frac{e^{-\mu|\text{\textbf{r}}_{k}-
\text{\textbf{r}}'|}}{|\text{\textbf{r}}_{k}-\text{\textbf{r}}'|}\nn\\
&\equiv& \frac{1}{18} +S_N
-\frac{\mu^2}{72\pi}\left[\mathcal{I}_1(r<R)+\frac{\mu}{2}\mathcal{I}_2(r<R)-\frac{\mu^2}{4\pi}\mathcal{I}_3(r<R)\right]\nn\\
&+& \frac{\mu^4}{288\pi^2\overline{\rho}}\sum\limits_{k=1}^{N}m_k
\left[
\frac{ e^{-\mu |\text{\textbf{r}}-\text{\textbf{r}}_{k} |}}{|\text{\textbf{r}}-\text{\textbf{r}}_{k}|}\ \int\limits_{r'>R} d{\bf r}'\
\frac{e^{-\mu|\text{\textbf{r}}'-\text{\textbf{r}}_k|}}{|\text{\textbf{r}}'-\text{\textbf{r}}_k|}
+\int\limits_{r'>R} d{\bf r}' \ \frac{ e^{-\mu |\text{\textbf{r}}-\text{\textbf{r}}' |}}{|\text{\textbf{r}}- \text{\textbf{r}}'|}\frac{e^{-\mu|\text{\textbf{r}}'-
\text{\textbf{r}}_k|}}{|\text{\textbf{r}}'-\text{\textbf{r}}_k|} \right]\, . \ea
Here the introduced term $S_N$ incorporates pure sums. The integrals $\mathcal{I}_1(r<R)$, $\mathcal{I}_2(r<R)$ and $\mathcal{I}_3(r<R)$ are given by
Eqs.~\rf{b.6}, \rf{b.8} and \rf{b.10}, respectively. In the last line of \rf{4.4} $\mathbf{r}_k$ are the radius-vectors of discrete particles inside the sphere.
To evaluate the expression in this line, we suppose that the discrete masses are concentrated in the central part of the sphere, i.e. $r_k \ll R \, \Rightarrow \,
r_k \ll r'$. Hence,
\ba{4.5} &{}&\frac{ e^{-\mu |\text{\textbf{r}}-\text{\textbf{r}}_{k} |}}{|\text{\textbf{r}}-\text{\textbf{r}}_{k}|}\ \int\limits_{r'>R} d{\bf r}'\
\frac{e^{-\mu|\text{\textbf{r}}'-\text{\textbf{r}}_k|}}{|\text{\textbf{r}}'-\text{\textbf{r}}_k|} +\int\limits_{r'>R} d{\bf r}' \ \frac{ e^{-\mu
|\text{\textbf{r}}-\text{\textbf{r}}' |}}{|\text{\textbf{r}}- \text{\textbf{r}}'|}\frac{e^{-\mu|\text{\textbf{r}}'-
\text{\textbf{r}}_k|}}{|\text{\textbf{r}}'-\text{\textbf{r}}_k|}\nn\\
&\approx& \frac{ e^{-\mu |\text{\textbf{r}}-\text{\textbf{r}}_{k} |}}{|\text{\textbf{r}}-\text{\textbf{r}}_{k}|}\int\limits_{r'>R} d{\bf r}'\ \frac{e^{-\mu
r'}}{r'}+\int\limits_{r'>R} d{\bf r}'\ \frac{ e^{-\mu |\text{\textbf{r}}- \text{\textbf{r}}' |}}{|\text{\textbf{r}}-\text{\textbf{r}}'|}\frac{e^{-\mu
r'}}{r'}\equiv \frac{ e^{-\mu |\text{\textbf{r}}-\text{\textbf{r}}_{k} |}}{|\text{\textbf{r}}-\text{\textbf{r}}_{k}|} \mathcal{I}_4 +\mathcal{I}_5(r<R)\, , \ea
where the integrals $\mathcal{I}_4$ and $\mathcal{I}_5(r<R)$ are given by Eqs.~\rf{b.11} and \rf{b.13}, respectively. Substituting this expression into \rf{4.4}
and taking into account the integrals \rf{b.6}, \rf{b.8}, \rf{b.10}, \rf{b.11} and \rf{b.13}, we get
\ba{4.6} &{}&\chi=
\frac{1}{18}+\frac{1}{36}e^{-\mu R}\left[\vphantom{e^{e^R}}\left(1+\mu R\right)\cosh(\mu r)\right.\nn\\
&-& \left.
\Big(3+3\mu R+\mu^2 R^2+\mu Re^{-\mu R} \cosh (\mu R) -e^{-\mu R}\sinh (\mu R)\Big)\frac{\sinh(\mu r)}{\mu r}\right] \nn\\
&+&\frac{\mu^2}{72\pi \bar{\rho}}\sum\limits_{k=1}^{N}m_k\frac{ e^{-\mu |\text{\textbf{r}}-
\text{\textbf{r}}_{k} |}}{|\text{\textbf{r}}-\text{\textbf{r}}_{k}|}
\Big[ \left(1+\mu R\right)e^{-\mu R}-1\Big]+\frac{\mu^3}{144\pi \bar{\rho}}e^{-2\mu R}\frac{\sinh (\mu r)}{\mu r}\sum\limits_{k=1}^{N}m_{k}\nn\\
&-&\frac{\mu^3}{144\pi \bar{\rho}}\sum\limits_{k=1}^{N}m_k\ e^{-\mu |\text{\textbf{r}}-\text{\textbf{r}}_k|}+\frac{\mu^4}{288\pi^2
\bar{\rho}^2}{\sum\limits_{k,k'=1}^{N}} {\vphantom{\int}}' m_k m_{k'}\ \frac{ e^{-\mu |\text{\textbf{r}}-\text{\textbf{r}}_{k} |}}{|\text{\textbf{r}}-
\text{\textbf{r}}_{k}|} \frac{e^{-\mu|\text{\textbf{r}}_{k'}-\text{\textbf{r}}_k|}}{|\text{\textbf{r}}_{k'}-\text{\textbf{r}}_k|}\, . \ea
In the limit $\mu R \to 0 \ \Rightarrow\ \mu r \to 0,\ \mu r_k \to 0$ we obtain
\be{4.7} \chi\approx \frac{\kappa c^2}{16 \pi a}\sum\limits_{k=1}^{N}\frac{m_k}{|\text{\textbf{r}}-\text{\textbf{r}}_{k}|}\frac{\kappa c^2}{8 \pi a}
\sum_{k'=1}^{N}\frac{m_{k'}}{|\text{\textbf{r}}_{k'}-\text{\textbf{r}}_k|} = -\frac{\kappa c^2}{16 \pi
a}\sum\limits_{k=1}^{N}\frac{m_k}{|\text{\textbf{r}}-\text{\textbf{r}}_{k}|}\left.\Phi_{N}\right|_{\mathbf{r}=\mathbf{r}_k}\, . \ee

Therefore, taking into account Eqs.~\rf{3.9} and \rf{4.3}, the potential $\Psi^{(2)}(\eta,\mathbf{r})$ tends to
\be{4.8} \Psi_N^{(2)}= -\frac{3}{4}\Phi_N^2 -\frac{\kappa c^2}{16 \pi
a}\sum\limits_{k=1}^{N}\frac{m_k}{|\text{\textbf{r}}-\text{\textbf{r}}_{k}|}\left.\Phi_{N}\right|_{\mathbf{r}=\mathbf{r}_k}\, ,\ee
in full agreement with the formula (3.73) in \cite{Eingorn-second}. This expression clearly demonstrates that $\Psi^{(2)}$ is really of the second order of
smallness.

To investigate the same limit of the second-order potential $\Phi^{(2)}(\eta,\mathbf{r})$, we need to substitute \rf{4.8} as well as the limit of $Q^{(S)}$ into
Eq.~\rf{3.26}.
One can show that
\be{4.9} Q_N^{(S)} =\frac{7}{4a^2}\Phi_N ^2 -\frac{\kappa c^2}{16 \pi
a^3}\sum\limits_{k=1}^{N}\frac{m_k}{|\text{\textbf{r}}-\text{\textbf{r}}_{k}|}\left.\Phi_{N}\right|_{\mathbf{r}=\mathbf{r}_k} +\frac{3 \kappa c^2}{16\pi a^3}
\sum\limits_{k=1}^{N}m_k \left(\frac{(\text{\textbf{r}}-\text{\textbf{r}}_k) \left.\nabla\Phi_N\right|_{\mathbf{r}=\mathbf{r}_k}}{|\text{\textbf{r}}-
\text{\textbf{r}}_k|}\right)\, . \ee
Consequently, we find
\be{4.10} \Phi_N^{(2)}= \Phi_N ^2 -\frac{\kappa c^2}{8 \pi
a}\sum\limits_{k=1}^{N}\frac{m_k}{|\text{\textbf{r}}-\text{\textbf{r}}_{k}|}\left.\Phi_{N}\right|_{\mathbf{r}=\mathbf{r}_k} +\frac{3 \kappa c^2}{16\pi a}
\sum\limits_{k=1}^{N}m_k \left(\frac{(\text{\textbf{r}}-\text{\textbf{r}}_k) \left.\nabla\Phi_N\right|_{\mathbf{r}=\mathbf{r}_k}}{|\text{\textbf{r}}-
\text{\textbf{r}}_k|}\right)\, . \ee
The expressions \rf{4.9} and \rf{4.10} agree with the formulas (3.74) and (3.75) in \cite{Eingorn-second}, respectively.


\

\subsection{Velocity-dependent second-order perturbations}

\

As is easily seen from Eqs.~\rf{5.12} and \rf{5.13}, the velocity-dependent second-order potentials in the post-Newtonian approximation read, respectively,
\be{4.11} \Psi_{(v)\,N}^{(2)}=-\frac{\kappa c^2}{16 \pi a}\sum\limits_{n}\frac{m_n \tilde{v}_n^2}{|{\bf r}-{\bf r}_{n}|} \ee
and
\be{4.12} \Phi^{(2)}_{(v)\,N}=-\frac{3\kappa c^2}{16 \pi a}\sum\limits_{n}\frac{m_n \tilde{v}_n^2}{|{\bf r}-{\bf r}_{n}|}+\frac{3\kappa c^2}{16 \pi
a}\sum\limits_{n}m_n\tilde{v}_n^\alpha \tilde{v}_n^\beta \frac{\partial ^2}{\partial x^\alpha \partial x^\beta} |{\bf r}-{\bf r}_{n}|\, . \ee




In the same approximation, the metric corrections $h_{00}$ and $h_{0\alpha}$ are presented, e.g., in the textbook \cite{Landau} (see Eqs.~(106.13) and (106.15),
respectively). Obviously, the sum of our expressions \rf{4.10} and \rf{4.12} should be in agreement with (106.13). The direct comparison shows that one-to-one
coincidence is absent. Indeed, the first two terms in the right-hand side of \rf{4.10} and the first term in the right-hand side of \rf{4.12} coincide (up to
evident redefinitions of spatial coordinates and peculiar velocities: $a{\bf r}\mapsto{\bf r}$, $a{\bf r}_n\mapsto{\bf r}_n$ and ${\bf\tilde v}_n\mapsto {\bf
v}_n/c$) with the corresponding terms in (106.13), divided by $2$, as it should be. However, each of Eqs.~\rf{4.10} and \rf{4.12} contain additional terms which
are absent in (106.13). As was emphasized in \cite{Eingorn-first} and \cite{Eingorn-second}, the reason for this noncoincidence with the Landau \& Lifshitz
formulas is the choice of different gauges here and in \cite{Landau}. To prove that the results are in accordance, let us consider an appropriate coordinate
transformation connecting our formulas and (106.13). Such a transformation was proposed in \cite{Eingorn-second}: $\eta \mapsto \eta -A(\eta,\mathbf{r})$. Then
$\Phi^{(2)} \mapsto \Phi^{(2)}+A'$. In this case the first-order vector perturbation is also transformed: $ \mathbf{B} \mapsto \mathbf{B} + \nabla A$. Hence, the
problem is to determine the function $A$ which provides the transition to the Landau \& Lifshitz formulas.

In the small-scale limit the first-order vector perturbation is presented in \cite{Eingorn-first} and \cite{Eingorn-second}. Its comparison with Eq.~(106.15) in
\cite{Landau} gives (see Eq.~(3.80) in \cite{Eingorn-second}):
\be{4.13} \frac{\partial A}{\partial x^\gamma}=\frac{3\kappa c^2}{16 \pi a}\sum\limits_{n}\left[\frac{m_n\tilde{v}_n^\gamma }{|{\bf r}-{\bf r}_{n}|}-\frac{m_n
\left[\tilde{v}_n^\beta (x^\beta-x^\beta_{n})\right]}{|{\bf r}-{\bf r}_{n}|^3} (x^\gamma-x^\gamma_{n})\right]\, . \ee
The time derivative of this expression reads
\ba{4.14} \left(\frac{\partial A}{\partial x^\gamma}\right)'&=& -\frac{3\kappa c^2}{16\pi a}\sum\limits_{n}\frac{m_n}{|{\bf r}-{\bf
r}_n|}\left\{\left.\left(\frac{\partial \Phi_N}{\partial x^\gamma}\right)\right|_{{\bf r} ={\bf
r}_n}-\left[(\text{\textbf{r}}-\text{\textbf{r}}_n)\left.\nabla\Phi_N\right|_{\mathbf{r}=\mathbf{r}_n}\right]
\frac{(x^{\gamma}-x^{\gamma}_n)}{|{\bf r}-{\bf r}_n|^2}\right\}\nn\\
&+&\frac{3\kappa c^2}{16 \pi a}\sum\limits_{n}m_n\tilde{v}_n^\alpha \tilde{v}_n^\beta \left[\frac{\delta^{\beta\gamma} (x^\alpha-x^\alpha_{n})}{|{\bf r}-{\bf
r}_n|^3}+\frac{\delta^{\alpha\gamma}(x^\beta-x^\beta_{n})+
\delta^{\alpha \beta}(x^\gamma-x^\gamma_{n})}{|{\bf r}-{\bf r}_n|^3}\right.\nn\\
&-& \left.\frac{3}{|{\bf r}-{\bf r}_n|^5}(x^\alpha-x^\alpha_{n})(x^\beta-x^\beta_{n})(x^\gamma-x^\gamma_{n})\right]\, . \ea
The first line of this formula is the time derivative of \rf{4.13} with fixed positions $\mathbf{r}_n$ (i.e. when the time derivatives of $\mathbf{r}_n$ are not
taken into account). This line was determined previously in \cite{Eingorn-second}. The second and third lines represent the time derivative of \rf{4.13} with
fixed velocities $\mathbf{\tilde{v}}_n$. These two lines are novel.

On the other hand, the direct comparison of the formula (106.13) with Eqs. \rf{4.10} and \rf{4.12} shows  that
\be{4.15} A' = -\frac{3 \kappa c^2}{16\pi a} \sum\limits_{n}m_n
\left(\frac{(\text{\textbf{r}}-\text{\textbf{r}}_n)\left.\nabla\Phi_N\right|_{\mathbf{r}=\mathbf{r}_n}}{|\text{\textbf{r}}- \text{\textbf{r}}_n|}\right)
-\frac{3\kappa c^2}{16 \pi a}\sum\limits_{n}m_n\tilde{v}_n^\alpha \tilde{v}_n^\beta \frac{\partial ^2}{\partial x^\alpha
\partial x^\beta}|{\bf r}-{\bf r}_n|\, .
\ee
It is not difficult to verify that the gradient of \rf{4.15} exactly coincides with \rf{4.14}. Therefore, we have proved that the sought-for function $A$ exists
and the agreement with the Landau \& Lifshitz formulas is achieved.


\section{Conclusion}
\label{Sec6}

\setcounter{equation}{0}

In the present paper we have studied the second-order scalar perturbations for the $\Lambda$CDM cosmological model within the cosmic screening approach. We have
found the analytic expressions for both the velocity-independent and velocity-dependent perturbations (see Eqs.~\rf{3.9}, \rf{3.26}, \rf{5.12}, \rf{5.13}). We
have demonstrated that the Yukawa screening effect, being inherent in the first-order metric corrections (see, e.g., \cite{Eingorn-first,EKZ1,EKZ2}), affects the
second-order ones as well. In addition, it has been shown that the obtained expressions for the second-order perturbations are reduced to the known post-Newtonian
formulas \cite{Eingorn-second,Landau} at distances much smaller than the screening length. With respect to the velocity-independent perturbations, such
correspondence was already observed in \cite{Eingorn-second}. However, in the case of the velocity-dependent perturbations this is a novel result.

In the era of precision cosmology, the derived analytic formulas play an important role since they enable revealing the contributions of the second-order metric
corrections to the observed physical quantities and effects (including the luminosity-redshift relation, gravitational lensing, baryon acoustic oscillations). The
analytic expressions for the second-order perturbations give an opportunity to estimate the backreaction and determine how strong the backreaction is and to what
extent we may use the standard FLRW metric as a background one. Obviously, the perturbative approach is robust if the second-order corrections are much smaller
than the first-order ones. If this is the case, it is usually enough to be limited to the first order. However, how can we know it from numerical simulations?
Performing a numerical simulation, which takes into account the second-order perturbations, is not an easy task. Instead of such a complicated procedure, we
suggest the following test. The simulation can be performed on the basis of the first-order approach. After that one can calculate the first-order perturbation
$\Phi$ at a number of points. On the other hand, using our analytic expressions, one can calculate the second-order perturbations $\Phi^{(2)}$ and $\Psi^{(2)}$ at
the same points. Then, if $\Phi \gg \Phi^{(2)}, \Psi^{(2)}$, the perturbative scheme is robust, and the backreaction is apparently negligible. Otherwise the
backreaction should be certainly taken into account.

\section*{Acknowledgements}

We would like to thank the Referee for the valuable comments which have considerably improved the presentation and discussion of the obtained results.

\appendix
\renewcommand{\theequation}{A.\arabic{equation}}

\section{Integrals}

\setcounter{equation}{0}

In this appendix we present the integrals used for our calculations. First, we list three integrals where integration over the radial coordinate runs from zero to
infinity:
\ba{b.1} I_1&=&\int d \text{\textbf{r}}' \frac{ e^{-a|\text{\textbf{r}}-\text{\textbf{r}}'|/\lambda}}{|\text{\textbf{r}}-\text{\textbf{r}}'|}
=4\pi \frac{\lambda ^2}{a^2}\, ;\\
I_2&=&\int d \text{\textbf{r}}' \frac{ e^{-a|\text{\textbf{r}}-\text{\textbf{r}}'|/\lambda}}{|\text{\textbf{r}}-
\text{\textbf{r}}'|}\frac{e^{-a|\text{\textbf{r}}'- \text{\textbf{r}}_k|/\lambda}}{|\text{\textbf{r}}'-\text{\textbf{r}}_k|}
=\frac{2\pi\lambda}{a}e^{-a|\text{\textbf{r}}-\text{\textbf{r}}_k|/\lambda}\label{b.2}\, ;\\
I_3&=&\int\frac{ d\text{\textbf{r}}'}{|\text{\textbf{r}}-\text{\textbf{r}}'|}\frac{e^{ -\mu |\text{\textbf{r}}'-
\text{\textbf{r}}_k|}}{|\text{\textbf{r}}'-\text{\textbf{r}}_k|}= \frac{4\pi}{\mu^2} \frac{1-e^{-\mu
|\text{\textbf{r}}-\text{\textbf{r}}_k|}}{|\text{\textbf{r}}-\text{\textbf{r}}_k|} \label{b.3}\, . \ea

Now we consider the integrals which correspond to the model described in section~\ref{Sec5}. In this case integration over the radial coordinate runs from the
radius $R$ of the sphere to infinity:
\ba{b.4} \mathcal{I}_1(r) &=& \int\limits_{r'>R} d{\bf r}' \frac{ e^{-\mu |\text{\textbf{r}}-\text{\textbf{r}}'|}}{|\text{\textbf{r}}-\text{\textbf{r}}'|}
=-\frac{2\pi}{\mu r} \int_R^{\infty} dr' r'  \left(e^{-\mu |r+r'|}-e^{-\mu |r-r'|}\right)
\, ,\\
\mathcal{I}_1(r<R) &=& \frac{4\pi }{\mu^2}e^{-\mu R}(1+\mu R)\frac{\sinh (\mu r)}{\mu r} \label{b.6}\, ;\ea
\ba{b.7} \mathcal{I}_2(r) &=& \int\limits_{r'>R} d{\bf r}'  e^{-\mu |\text{\textbf{r}}-\text{\textbf{r}}'|}\, ,\\
\mathcal{I}_2(r<R) &=& \frac{4\pi}{\mu^3}e^{-\mu R} \left[\frac{\sinh(\mu r)}{\mu r} \left(3+3\mu R+\mu ^2R^2\right)- \cosh(\mu r)\left(1+\mu R\right)\right]
\label{b.8}\, ;\ea
\ba{b.9} \mathcal{I}_3(r) &=& \int\limits_{r'>R} d{\bf r}'\int\limits_{r''>R} d{\bf r}'' \frac{ e^{-\mu
|\text{\textbf{r}}-\text{\textbf{r}}'|}}{|\text{\textbf{r}}- \text{\textbf{r}}'|}\frac{ e^{-\mu
|\text{\textbf{r}}''-\text{\textbf{r}}'|}}{|\text{\textbf{r}}''-\text{\textbf{r}}'|}\, ,\\
\mathcal{I}_3(r<R) &=& \frac{8\pi^2}{\mu^4} \frac{\sinh (\mu r)}{\mu r}e^{-\mu R}\left[2(1+\mu R)+\left(-\mu R \cosh (\mu R)+ \sinh (\mu R)\right)e^{-\mu
R}\right] \label{b.10}\, ;\ea
\be{b.11} \mathcal{I}_4 = \int\limits_{r'>R} d{\bf r}'\  \frac{e^{-\mu r'}}{r'} = \frac{4\pi }{\mu^2}\left(1+\mu R\right)e^{-\mu R}\, ;\ee
\ba{b.12} \mathcal{I}_5(r) &=& \int\limits_{r'>R} d{\bf r}'\ \frac{ e^{-\mu |\text{\textbf{r}}-\text{\textbf{r}}' |}}{|\text{\textbf{r}}-
\text{\textbf{r}}'|}\frac{e^{-\mu r'}}{r'}\, ,\\
\mathcal{I}_5(r<R) &=& \frac{2\pi}{\mu}e^{-2\mu R}\frac{\sinh (\mu r)}{\mu r} \label{b.13}\, . \ea







\end{document}